\title{Investigating the source of Planck-detected AME: high resolution observations at 15 GHz}
\author{
        Yvette~C.~Perrott\\
	Astrophysics Group, Cavendish Laboratory, \\
        19 J.~J.~Thomson Avenue, Cambridge CB3 0HE, \underline{UK}
	\and
	Anna M.~M.~Scaife$^{1}$, Natasha Hurley-Walker$^{2}$ \& Keith J. B. Grainge$^{3}$.\\
	$^{1}$ Physics \& Astronomy, University of Southampton, \\
	Highfield, Southampton, SO17 1BJ, \underline{UK}; \\
	$^{2}$ International Centre for Radio Astronomy Research,\\ 
	Curtin Institute of Radio Astronomy, 1 Turner Avenue, \\
	Technology Park, Bentley, WA 6845, \underline{Australia}; \\
	$^{3}$ Astrophysics Group, Cavendish Laboratory, \\
        19 J.~J.~Thomson Avenue, Cambridge CB3 0HE, \underline{UK} \\
}

\date{\today}

\documentclass[letterpaper,11pt]{article}
\usepackage{aas_macros}
\usepackage{epsfig}
\usepackage{color, psfrag}
\newcommand{\todo}[1]{}
\begin{document}

\maketitle

\begin{abstract}
The \emph{Planck} 28.5\,GHz maps were searched for potential Anomalous Microwave Emission (AME) regions on the scale of $\sim3^{\circ}$ or smaller, and several new regions of interest were selected.  Ancillary data at both lower and higher frequencies were used to construct spectral energy distributions (SEDs), which seem to confirm an excess consistent with spinning dust models.  Here we present higher resolution observations of two of these new regions with the Arcminute Microkelvin Imager Small Array (AMI SA) between 14 and 18\,GHz to test for the presence of a compact ($\sim$10\,arcmin or smaller) component.  For AME-G107.1+5.2, dominated by the {\sc Hii} region S140, we find evidence for the characteristic rising spectrum associated with the either the spinning dust mechanism for AME or an ultra/hyper-compact \textsc{Hii} region across the AMI frequency band, however for AME-G173.6+2.8 we find no evidence for AME on scales of $\sim 2-10$\,arcmin.
\end{abstract}


\maketitle

\section{Introduction}

The \emph{Planck} satellite (\cite{2010A&A...520A...1T}, \cite{2011A&A...536A...1P}) observes the sky in nine frequency bands, covering a range from 30 to 857\,GHz.  Its wide frequency range potentially allows the detection of AME since the high frequency data above 100\,GHz can be used to constrain the thermal emission, while the lower frequency data is close to the theoretical peak of the spinning dust emission.  When combined with ancillary data at lower frequencies, the spectra of AME regions can be accurately determined and used to probe their properties on large ($\sim 1$\,degreee) scales.  The \emph{Planck} maps were used to detect several new potential AME regions \cite{2011A&A...536A..20P} by subtracting a spatial model of known emission mechanisms (synchrotron, free-free and thermal dust) extrapolated from observational or theoretical predictions.  Two, AME-G173.6+2.8 and AME-G107.1+5.2, were selected and ancillary data were used at both higher and lower frequencies to construct SEDs, which contain suggestions of AME consistent with spinning dust emission.

The Arcminute Microkelvin Imager Small Array (AMI-SA) is a radio interferometer situated near Cambridge, UK.  Primarily an SZ survey instrument, the AMI-SA is specifically designed to have high sensitivity to low-surface-brightness emission on scales of $2-$10\,arcmin.  It operates between 14 and 18\,GHz, close in frequency to the theoretical peak of the spinning dust emission, the position of which varies between $10-50$\,GHz depending on grain size composition. The AMI-SA has previously been used both to identify and to characterize spinning dust regions in multiple Galactic (e.g. \cite{2009MNRAS.394L..46A}, \cite{2010MNRAS.403L..46S}) and extra-galactic (\cite{ngc6946}) sources. The higher angular resolution and lower frequency coverage of the AMI-SA make it a highly complementary instrument to the \emph{Planck} satellite for studies of AME. The synthesised beam of the SA, which is an effective measure of the resolution, is $\simeq$2\,arcmin at FWHM, while the \emph{Planck} maps and ancillary data were smoothed to a common resolution of $\simeq1^{\circ}$ for AME detection.

\section{Observations and Data Reduction}

The AMI SA is situated at the Mullard Radio Astronomy Observatory, Cambridge \cite{2008MNRAS.391.1545Z}. It consists of ten 3.7-m-diameter dishes with a baseline range of $\simeq 5$--20\,m and observes in the band 12--18\,GHz with eight 0.75-GHz bandwidth channels.  In practice, the lowest two frequency channels are unused due to a low response in this frequency range, and interference from geostationary satellites.

The two sources listed in Table~\ref{tab:srclist} were observed from 2011 Jan 21 $-$ Feb 09, for $\simeq$5\,hours each.

\begin{table}[h!]
\begin{center}
\caption{The \emph{Planck} AME candidates observed with the AMI SA, and the sources used as phase calibrators.\label{tab:srclist}}
\begin{tabular}{lcccc}
\hline \hline
\emph{Planck} ID  & Other & RA  & $\delta$ & Phase  \\
           & names & (J2000) & (J2000) & calibrator\\  
\hline 
AME-G173.6$+$2.8 & S235 & 05 41 06.0 & +35 50 00 & J0555+3948 \\
AME-G107.1$+$5.2 & S140 & 22 19 18.1 & +63 18 49 & J2125+6423 \\ 
\hline 
\end{tabular}
\end{center}
\end{table}


Data reduction was performed using the local software tool \textsc{reduce}, which flags interference, shadowing and hardware errors, applies phase and amplitude calibrations and Fourier transforms the correlator data to synthesize the frequency channels, before output to disk in $uv$ \textsc{fits} format.  Flux calibration was performed using short observations of 3C48, 3C286 or 3C147 near the beginning and end of each run.  The assumed flux densities for 3C286 were converted from Very Large Array total-intensity measurements provided by R. Perley (private communication), and are consistent with the Rudy et al.(1987) \cite{1987Icar...71..159R} model of Mars transferred on to absolute scale, using results from the \emph{Wilkinson Microwave Anisotropy Probe}.  The assumed flux densities for 3C48 and 3C147 are based on long-term monitoring with the AMI SA using 3C286 for flux calibration (see Table~\ref{tab:Fluxes-of-3C286}).  A correction for changing airmass is also applied using a noise-injection system, the `rain gauge'.

\begin{table}
\centering
\caption{Assumed I~+~Q flux densities of 3C286, 3C48 and 3C147.} \label{tab:Fluxes-of-3C286}
\bigskip
\begin{tabular}{ccccc}
\hline \hline
 Channel & $\bar{\nu}$/GHz & $S^{\rm 3C286}$/Jy & $S^{\rm 3C48}$/Jy & $S^{\rm 3C147}$/Jy \phantom{$S^{\rm{X^{X}}}$} \\ 
\hline 
 3 & 13.88 & 3.74 & 1.89 & 2.72 \\
 4 & 14.63 & 3.60 & 1.78 & 2.58 \\
 5 & 15.38 & 3.47 & 1.68 & 2.45 \\
 6 & 16.13 & 3.35 & 1.60 & 2.34 \\
 7 & 16.88 & 3.24 & 1.52 & 2.23 \\
 8 & 17.63 & 3.14 & 1.45 & 2.13 \\ 
\hline 
\end{tabular}
\end{table}

Bright, nearby point sources selected from the Very Long Baseline Array Calibrator Survey \cite{1998ASPC..144..155P} were observed during each observation at hourly intervals for phase calibration purposes (see Table~\ref{tab:srclist} for phase calibrators used for the AMI SA observations).  The reduced visibility data were imaged using \textsc{aips}\footnote{\texttt{http://aips.nrao.edu/}}, from the individual channel datasets (for channels 3 to 8 inclusive), as well as from the combined channels at a central frequency of 15.75\,GHz.  Gaussians were fitted to the sources detected at $>5\sigma$ in the maps using the \textsc{aips} task \textsc{jmfit}.  Errors on AMI SA flux density values were estimated by adding in quadrature the error output from \textsc{jmfit}, which folds in an estimate of the r.m.s. map noise and the error associated with the Gaussian fit, and the error on flux calibration (including rain-gauge correction) of $\simeq 5$ per cent of the integrated flux density.

\subsection{Matching spatial scales}

\begin{figure}
  \begin{center}
	\centerline{
%
%
\begin{psfrags}%
\psfragscanon%
\Large
\psfrag{s01}[t][t]{\color[rgb]{0,0,0}\setlength{\tabcolsep}{0pt}\begin{tabular}{c}CGPS sampled visibility amplitude (Jy)\end{tabular}}%
\psfrag{s02}[b][b]{\color[rgb]{0,0,0}\setlength{\tabcolsep}{0pt}\begin{tabular}{c}AMI visibility amplitude (Jy)\end{tabular}}%
%
\psfrag{x01}[t][t]{0}%
\psfrag{x02}[t][t]{0.1}%
\psfrag{x03}[t][t]{0.2}%
\psfrag{x04}[t][t]{0.3}%
\psfrag{x05}[t][t]{0.4}%
\psfrag{x06}[t][t]{0.5}%
\psfrag{x07}[t][t]{0.6}%
\psfrag{x08}[t][t]{0.7}%
\psfrag{x09}[t][t]{0.8}%
\psfrag{x10}[t][t]{0.9}%
\psfrag{x11}[t][t]{1}%
\psfrag{x12}[t][t]{0}%
\psfrag{x13}[t][t]{1}%
\psfrag{x14}[t][t]{2}%
\psfrag{x15}[t][t]{3}%
%
\psfrag{v01}[r][r]{0}%
\psfrag{v02}[r][r]{0.1}%
\psfrag{v03}[r][r]{0.2}%
\psfrag{v04}[r][r]{0.3}%
\psfrag{v05}[r][r]{0.4}%
\psfrag{v06}[r][r]{0.5}%
\psfrag{v07}[r][r]{0.6}%
\psfrag{v08}[r][r]{0.7}%
\psfrag{v09}[r][r]{0.8}%
\psfrag{v10}[r][r]{0.9}%
\psfrag{v11}[r][r]{1}%
\psfrag{v12}[r][r]{0}%
\psfrag{v13}[r][r]{1}%
\psfrag{v14}[r][r]{2}%
\psfrag{v15}[r][r]{3}%
%
\resizebox{0.45\textwidth}{!}{\includegraphics{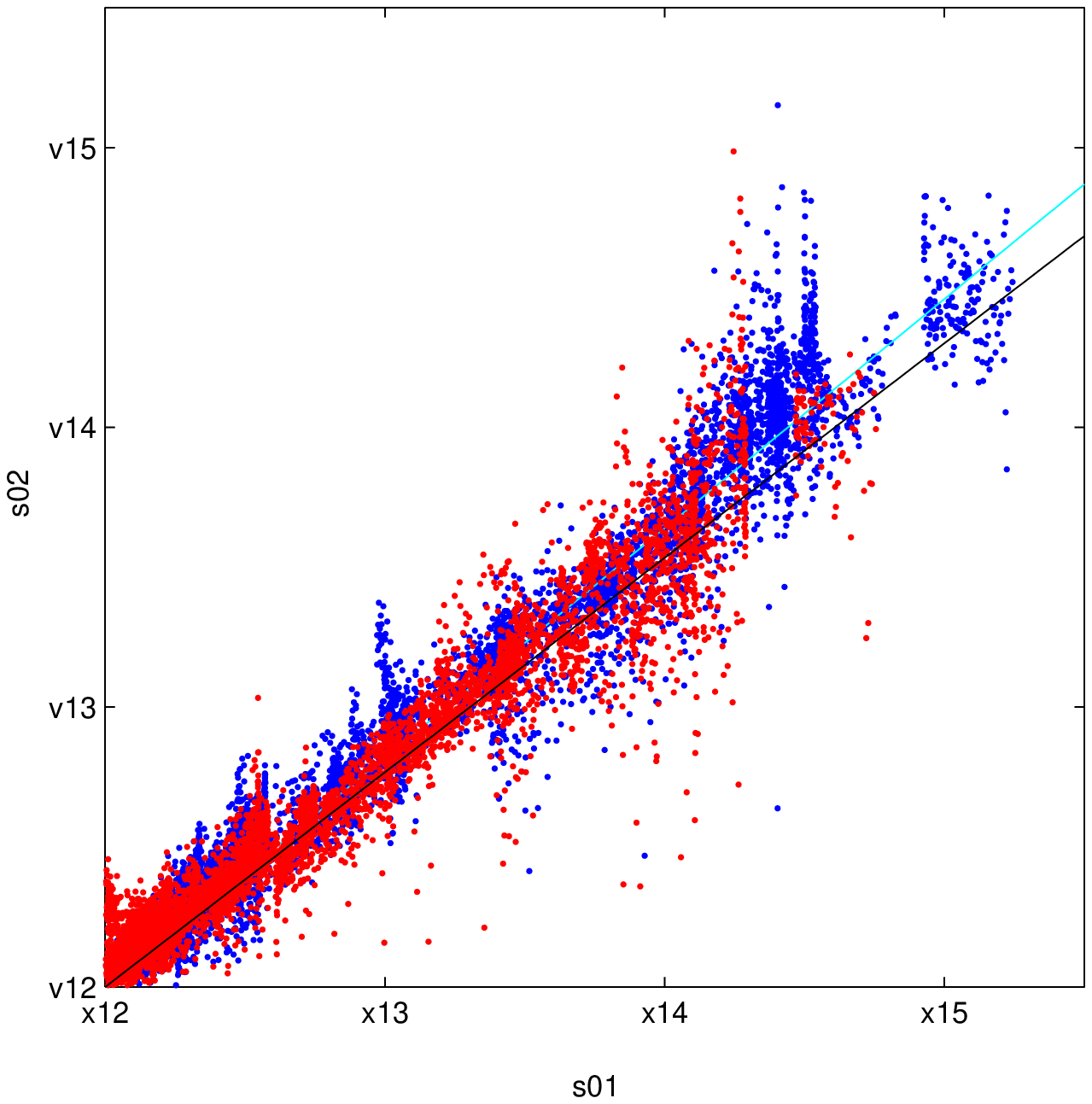}}%
\end{psfrags}%
%
\qquad
%
%
\begin{psfrags}%
\psfragscanon%
\Large
\psfrag{s01}[t][t]{\color[rgb]{0,0,0}\setlength{\tabcolsep}{0pt}\begin{tabular}{c}CGPS sampled visibility amplitude (Jy)\end{tabular}}%
\psfrag{s02}[b][b]{\color[rgb]{0,0,0}\setlength{\tabcolsep}{0pt}\begin{tabular}{c}AMI visibility amplitude (Jy)\end{tabular}}%
%
\psfrag{x01}[t][t]{0}%
\psfrag{x02}[t][t]{0.1}%
\psfrag{x03}[t][t]{0.2}%
\psfrag{x04}[t][t]{0.3}%
\psfrag{x05}[t][t]{0.4}%
\psfrag{x06}[t][t]{0.5}%
\psfrag{x07}[t][t]{0.6}%
\psfrag{x08}[t][t]{0.7}%
\psfrag{x09}[t][t]{0.8}%
\psfrag{x10}[t][t]{0.9}%
\psfrag{x11}[t][t]{1}%
\psfrag{x12}[t][t]{0}%
\psfrag{x13}[t][t]{0.1}%
\psfrag{x14}[t][t]{0.2}%
\psfrag{x15}[t][t]{0.3}%
\psfrag{x16}[t][t]{0.4}%
%
\psfrag{v01}[r][r]{0}%
\psfrag{v02}[r][r]{0.1}%
\psfrag{v03}[r][r]{0.2}%
\psfrag{v04}[r][r]{0.3}%
\psfrag{v05}[r][r]{0.4}%
\psfrag{v06}[r][r]{0.5}%
\psfrag{v07}[r][r]{0.6}%
\psfrag{v08}[r][r]{0.7}%
\psfrag{v09}[r][r]{0.8}%
\psfrag{v10}[r][r]{0.9}%
\psfrag{v11}[r][r]{1}%
\psfrag{v12}[r][r]{0}%
\psfrag{v13}[r][r]{0.2}%
\psfrag{v14}[r][r]{0.4}%
\psfrag{v15}[r][r]{0.6}%
\psfrag{v16}[r][r]{0.8}%
%
\resizebox{0.45\textwidth}{!}{\includegraphics{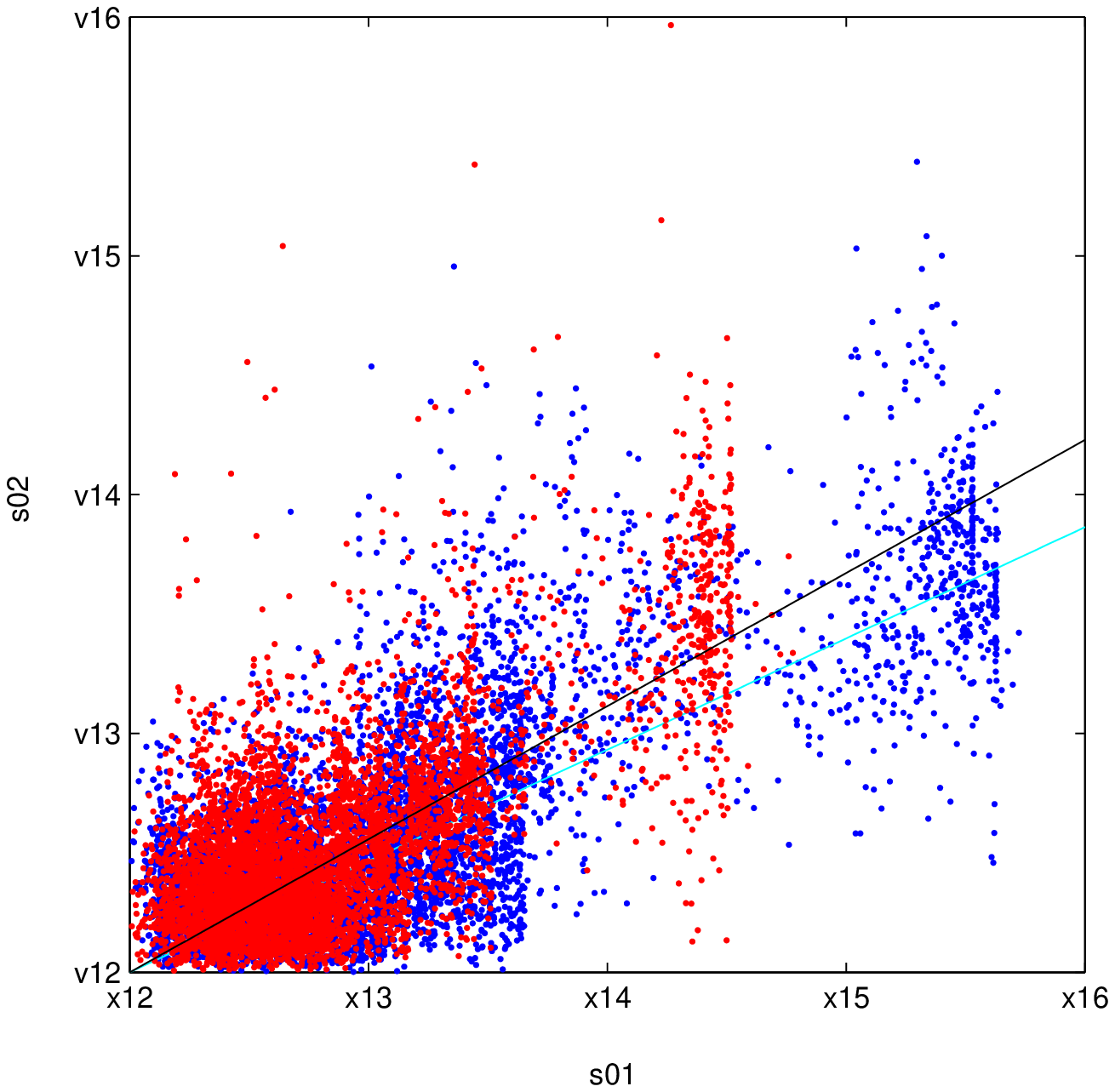}}%
\end{psfrags}%
%
}
	\centerline{(a) \hskip .45\textwidth (b)}
	\caption{(a) S235 and (b) S140. Visibilities sampled from CGPS 1.42\,GHz data are plotted on the x-axis, with the corresponding AMI observed visibilities on the y-axis for channels 3 (blue points) and 8 (red points).  The solid cyan and black lines show the respective ratios derived from fitting elliptical Gaussians to the maps, i.e. (integrated flux density from AMI channel map)/(integrated flux density from CGPS map sampled with corresponding AMI channel \textit{uv}-coverage).  There is a very good correlation between the sampled and true AMI visibilities, indicating that the morphology that AMI observes is very similar to that observed by CGPS.\label{fig:vis}}
  \end{center}
\end{figure}

Since the sources detected were extended to the AMI SA beam, flux loss corrections were calculated and applied by sampling Canadian Galactic Plane Survey (CGPS, \cite{2003AJ....125.3145T}) 1.4\,GHz total power maps with the \textit{uv}-coverage of the AMI observations, followed by mapping and fitting Gaussians to the sampled maps in the same manner as the AMI observations.  Fig.~\ref{fig:vis} shows the sampled visibilities compared to the AMI observed visibilities, and the ratios derived from fitting elliptical Gaussians to the maps.  As a consistency check, the flux loss corrections were also calculated by modeling the sources as Gaussians based on the size parameters derived from the AMI continuum maps.  The channel 3 flux density was taken as the reference, and channels 4 -- 8 and the CGPS 1.42\,GHz flux density were corrected to the channel 3 scale using the fitted flux loss percentages.

Lower frequency flux densities at 0.408, 2.7 and 4.85\,GHz are estimated from the lower resolution CGPS, Effelsberg and GB6 \cite{1996ApJS..103..427G} maps by convolving the CGPS 1.42\,GHz maps to the appropriate resolution, measuring a flux density on the convolved CGPS and lower resolution maps, and calculating a spectral index.  The lower resolution flux density is then normalised to the flux density estimated from the CGPS map sampled with AMI channel 3 \textit{uv}-coverage using the calculated spectral index.  These fluxes are however considered as upper limits since the amount of extra background flux density measured by the lower resolution maps is uncertain and not accounted for in the flux loss correction.

\section{Results}

AMI-SA maps of both sources are shown in Figs.~\ref{Fi:PSJ0541}~\&~\ref{Fi:PSJ2219}. The maps displayed are not corrected for attenuation due to the primary beam; the flux densities reported have been so corrected.  Where spectral indices $\alpha$ are quoted, the convention $S \propto \nu^{-\alpha}$ is used, where $S$ is flux density and $\nu$ is frequency.  Errors quoted are 1$\sigma$.

\subsection{AME-G173.6$+$2.8}

This region is centred on the \textsc{HII} region S235~\cite{1959ApJS....4..257S}, part of the FVW\,172.8+1.5 complex within Auriga. FVW\,172.8+1.5 is an {\sc Hii} complex composed of Sharpless {\sc Hii} regions and OB associations in the Perseus arm with individual objects organized along a large ($\sim7-4$\,deg) filamentary structure, the morphology of which has been compared to a bow tie \cite{2007MNRAS.379..289K}. Several {\sc Hii} regions within this complex have been the subject of previous study (e.g. \cite{1978A&A....63..325I}). The north-eastern regions: S231, S232, S233, and S235, are known to be associated with a giant molecular cloud at an adopted distance of 1.8\,kpc (\cite{1981ApJ...246..394E}; individual distances determined by spectrophotometry to each of the exciting stars within the {\sc Hii} regions range from 1.0 to 2.3\,kpc), and have systematic CO radial velocities of $-$18.1 to $-$23.0\,km\,s$^{-1}$ \cite{1982ApJS...49..183B}.  

It has recently been proposed that these four {\sc Hii} regions lie on the shell of a SNR \cite{2012AJ....143...75K}. This premise, rather than that of an expanding {\sc Hii} region, is supported by the fast expansion velocity of the shell and high kinetic energy associated with it. In addition the filamentary structures joining individual {\sc Hii} regions were shown to have non-thermal spectral contributions, inconsistent with an {\sc Hii} bubble.

\begin{figure}[ht]
  \begin{center}
    \includegraphics[bb=46 138 570 668, clip=,width=0.8\textwidth]{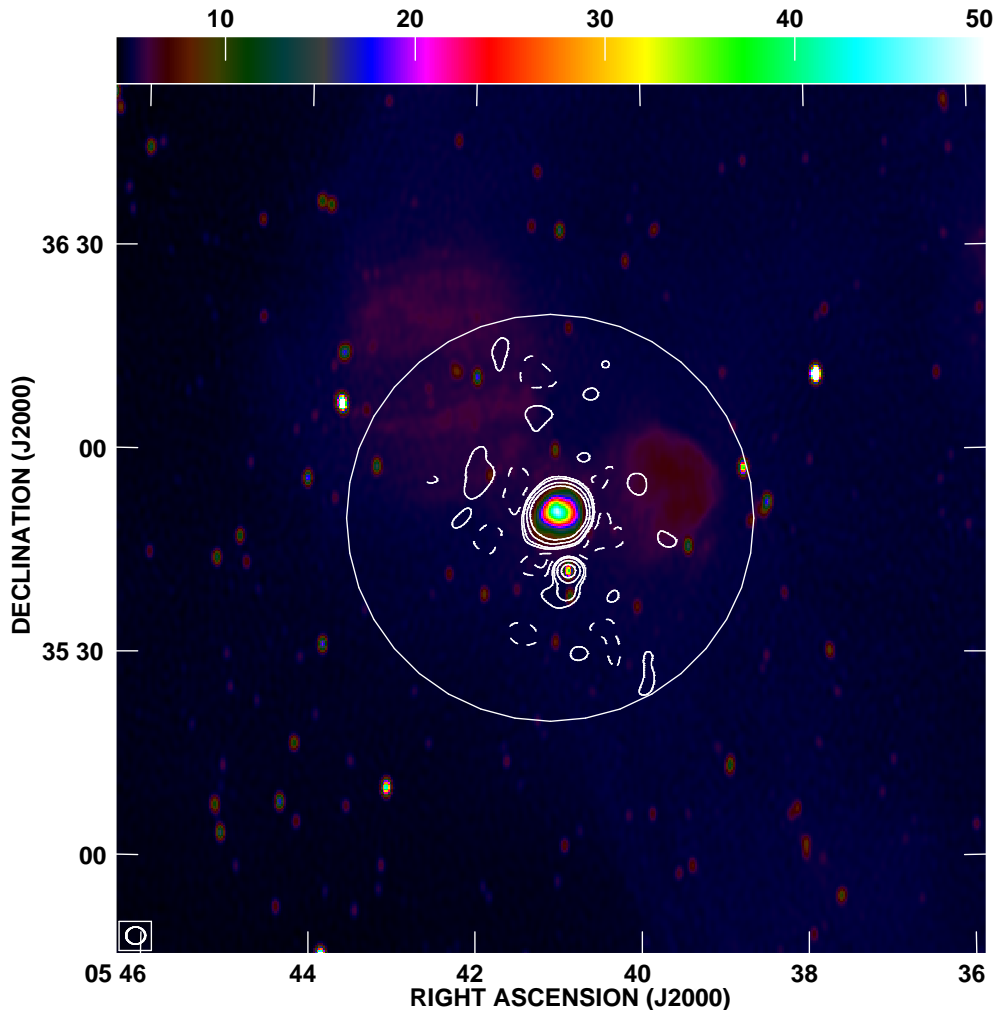}
    \caption{\textbf{AME-G173.6$+$2.8:} AMI SA data is shown as contours over CGPS 1.4\,GHz pseudo-colour, in brightness temperature.  The pseudo-colour scale is truncated at 50\,K ($\simeq$ peak brightness of S235) to show the low-surface-brightness diffuse structures.  AMI contours are at $\pm$5, 10, 25, and 50 $\sigma_{rms}$ on the AMI map (1.77\,mJy\,beam$^{-1}$).  The 1$^{\circ}$ \emph{Planck} smoothing scale is shown as the large circle and the AMI SA synthesised beam is shown in the bottom left corner. \label{Fi:PSJ0541}}
  \end{center}
\end{figure}

Fig.~\ref{Fi:PSJ0541} shows the AMI SA channel-averaged map and the corresponding CGPS 1.4\,GHz map.  The 1$^{\circ}$ smoothing resolution of the \emph{Planck} analysis is shown for comparison.  It can be seen that there are many components within the smoothing radius, both diffuse and compact, which could potentially contribute to the excess emission seen by \emph{Planck}.

A number of discrete sources are detected within the AMI-SA primary beam towards this region. The field is dominated by the north-south string of {\sc Hii} regions: S235, S235A (BFS46) and BFS47. BFS46 (S235A) was previously part of an AME study of {\sc Hii} regions \cite{2008MNRAS.385..809S} where it was found to exhibit no excess, having a spectrum consistent with  optically thin free-free emission with $\alpha=0.09\pm0.03$. This provides a useful check for any systematic offsets in individual channel calibration that may affect the spectrum of S235.  Although it lies away from the centre of the pointing where phase errors are expected to have greater effect, the flux densities derived from the AME-G173.6$+$2.8 map for BFS46 are consistent with those from the previous study.  


Here we also confirm that the spectrum of S235 is consistent with optically thin free-free emission.  After flux loss correction using the CGPS sampled maps, a spectral index of $\alpha=0.34\pm0.03$ is fitted to the AMI channel flux densities.  Using a simulated elliptical Gaussian with the deconvolved source size and position angle from the AMI continuum map as a consistency check gives a spectral index of $\alpha=0.57\pm0.03$, and it is likely that the difference indicates that this source is not well-modelled by a Gaussian.  Fig.~\ref{fig:vis}(a) shows the CGPS sampled visibilities compared to the AMI visibilities, and the correspondence derived from fitting to the maps.  It can be seen that the correlation between the visibilities is very good, indicating a good morphological correlation between 1.42 and 16\,GHz.

\begin{figure}
  \begin{center}
	\centerline{
%
%
\begin{psfrags}%
\psfragscanon%
\Large
\psfrag{s08}[t][t]{\color[rgb]{0,0,0}\setlength{\tabcolsep}{0pt}\begin{tabular}{c}Frequency (GHz)\end{tabular}}%
\psfrag{s09}[b][b]{\color[rgb]{0,0,0}\setlength{\tabcolsep}{0pt}\begin{tabular}{c}Flux density (Jy)\end{tabular}}%
%
\psfrag{x01}[t][t]{0}%
\psfrag{x02}[t][t]{0.1}%
\psfrag{x03}[t][t]{0.2}%
\psfrag{x04}[t][t]{0.3}%
\psfrag{x05}[t][t]{0.4}%
\psfrag{x06}[t][t]{0.5}%
\psfrag{x07}[t][t]{0.6}%
\psfrag{x08}[t][t]{0.7}%
\psfrag{x09}[t][t]{0.8}%
\psfrag{x10}[t][t]{0.9}%
\psfrag{x11}[t][t]{1}%
\psfrag{x12}[t][t]{17}%
\psfrag{x13}[t][t]{15}%
\psfrag{x14}[t][t]{0.5}%
\psfrag{x15}[t][t]{1}%
\psfrag{x16}[t][t]{5}%
\psfrag{x17}[t][t]{10}%
%
\psfrag{v01}[r][r]{0}%
\psfrag{v02}[r][r]{0.1}%
\psfrag{v03}[r][r]{0.2}%
\psfrag{v04}[r][r]{0.3}%
\psfrag{v05}[r][r]{0.4}%
\psfrag{v06}[r][r]{0.5}%
\psfrag{v07}[r][r]{0.6}%
\psfrag{v08}[r][r]{0.7}%
\psfrag{v09}[r][r]{0.8}%
\psfrag{v10}[r][r]{0.9}%
\psfrag{v11}[r][r]{1}%
\psfrag{v12}[r][r]{2.5}%
\psfrag{v13}[r][r]{2.7}%
\psfrag{v14}[r][r]{2.5}%
\psfrag{v15}[r][r]{3}%
\psfrag{v16}[r][r]{3.5}%
\psfrag{v17}[r][r]{4}%
%
\resizebox{0.469\textwidth}{!}{\includegraphics{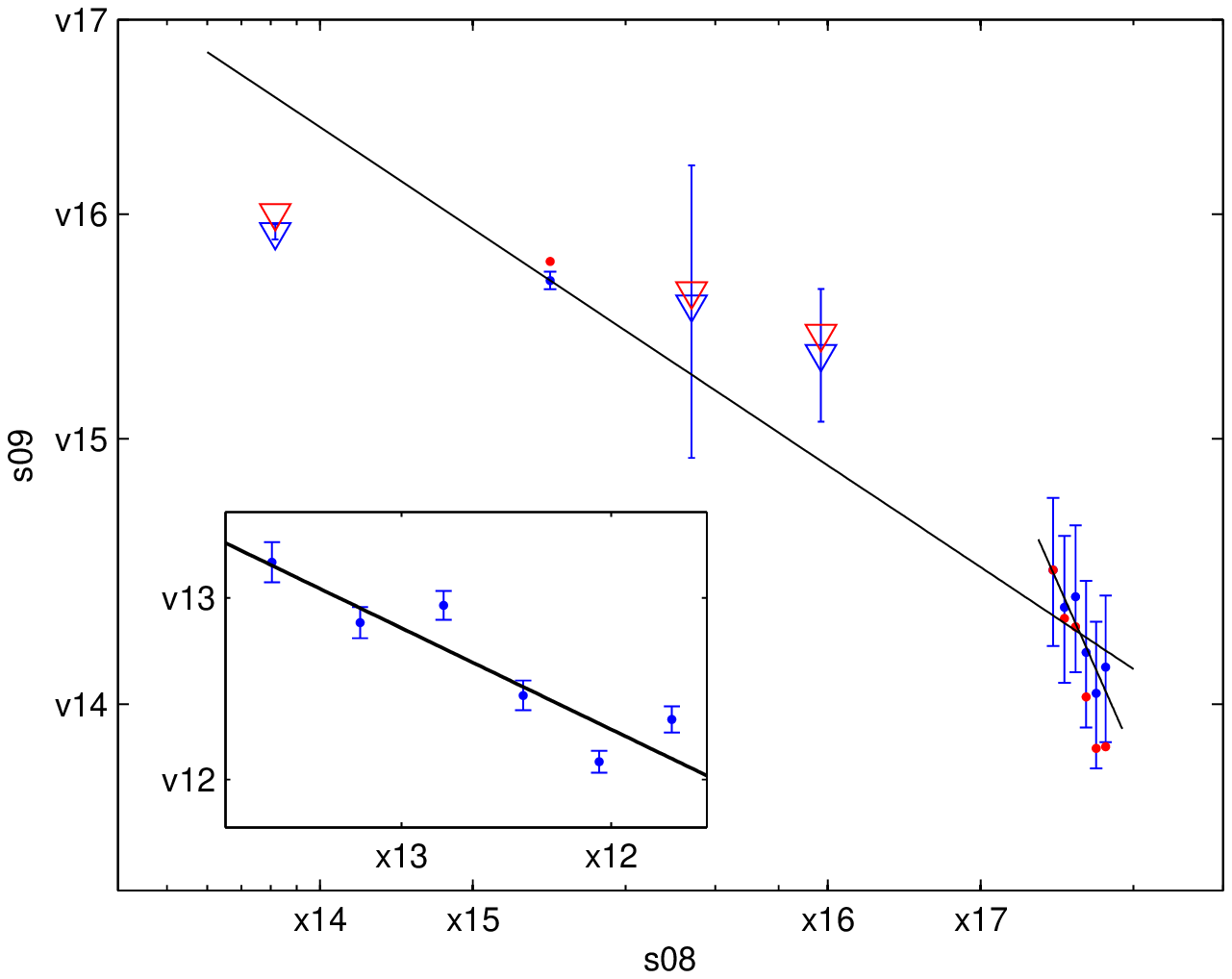}}%
\end{psfrags}%
%
\qquad\includegraphics[width=0.45\textwidth]{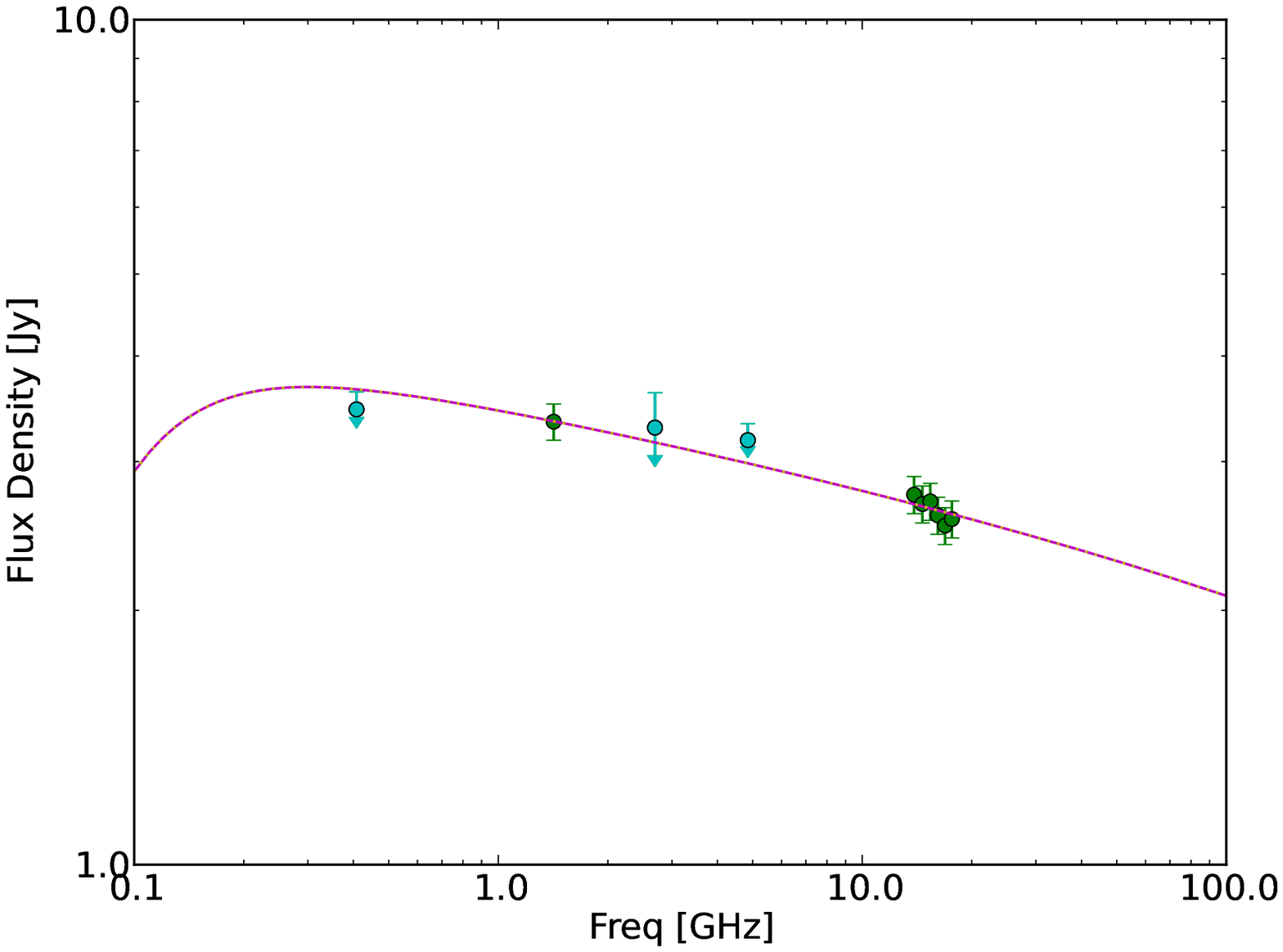}}
	\centerline{(a) \hskip .45\textwidth (b)}
    	\caption{SED for AME-G173.6$+$2.8, including flux densities at 0.408 and 1.42\,GHz from the CGPSE catalogue, 2.7\,GHz from Effelsberg, 4.85\,GHz from the GB6 catalogue corrected by the flux loss percentage of the AMI channel 3 observation, and the flux-loss-corrected AMI fluxes.  (a) shows flux densities corrected using the sampled CGPS map (blue), and using a simulated Gaussian source as a consistency check (red).  Lower resolution flux densities are shown as triangles as they are considered to be upper limits.  Also shown are the spectral index fits to CGPS 1.42\,GHz and AMI together, and to the blue AMI points separately. The inset plot is zoomed in to the AMI flux densities; here the errors plotted are the \textsc{jmfit} error estimates only (without calibration uncertainty).  (b) shows a free-free spectrum fit to the flux densities (see text for details).\label{Fi:PSJ0541_spec}}
  \end{center}
\end{figure}

Although a spectrum with an index of  $\alpha=0.34$ is steeper than the canonical free-free index of $\alpha=0.1$, such indices are expected for bremsstrahlung from cooler (heavy) plasma \cite{2003MNRAS.341..369D}. However, it is more likely in this instance that the steeper spectrum across the AMI band arises as a consequence of inexact flux-loss corrections due to differences in the structure of the source between 1.4\,GHz and the frequency range covered by AMI. Fitting to the CGPS 1.42\,GHz catalogue flux for S235 and the AMI channel fluxes gives a spectral index of $\alpha=0.101\pm0.008$, consistent with the radio recombination line measurements of Lockman \cite{1989ApJS...71..469L} who found $\Delta v = 22.3\pm2.4$\,km\,s$^{-1}$ for this region, indicative of optically thin plasma. Fig.~\ref{Fi:PSJ0541_spec}(a) shows the SED indicating the corrections for flux loss; it can be clearly seen that AMI sees no AME. Flux densities from the Effelsberg 11\,cm and GB6 4.85\,GHz surveys should be considered upper limits as the flux loss corrections calculated from the CGPS data fitted at a resolution of 1\,arcmin cannot account for any contribution from proximal low surface brightness emission which has contributed to the flux density of sources fitted at resolutions of 4.3 and 3\,arcmin, respectively. Indeed, when considered in conjunction with lower frequency data, Fig.~\ref{Fi:PSJ0541_spec}(b), we find that the AMI data is well-fitted by a free-free spectrum (\cite{1961AJ.....66R..50O}; see Scaife 2012, this volume). Fixing the electron temperature at $T_{\rm e}=8000$\,K, we find an emission measure of $EM=2.54\pm0.04\times 10^4$\,cm$^{-6}$\,pc, assuming a source with deconvolved size as determined from the AMI data ($\Omega_{\rm S235} = 232.6\times 208.9''$ P.A. = 104.5$^{\circ}$).

\subsection{AME-G107.1$+$5.2}

This region is centred on the \textsc{HII} region S140~\cite{1959ApJS....4..257S}. At a distance of 910\,pc, S140 is a bright-rimmed cloud that forms the interface between an {\sc Hii} region and the molecular cloud L1204 \cite{1974PDAO...14..283C}. S140 contains an IR cluster of at least three sources \cite{1979ApJ...232L..47B}. IRS1 is the progenitor of a large molecular outflow in the SE-NW direction with a reflection nebula associated to the blue-shifted SE lobe (\cite{1987ApJ...312..327H}; see also \cite{2007arXiv0711.4912H}). The activity of this outflow has caused significant disturbance in the surrounding area where a number of sweeping feather-like structures are visible in high angular resolution K-band images \cite{1986ApJ...311L..81F}. Perpendicular to the outflow a disk has been detected at cm wavelengths, confirmed by K-band polarimetric imaging (\cite{2006ApJ...649..856H}; \cite{2008ApJ...673L.175J}).

The AMI observation of this region is shown in Fig.~\ref{Fi:PSJ0541}. After flux loss correction, a spectral index of $\alpha=-0.75\pm0.05$ is fitted to the AMI channel fluxes.  The spectral index calculated using a simulated Gaussian source with size derived from the AMI channel-averaged map for flux loss correction is $\alpha=0.66\pm0.06$; however it can be seen that the source is highly non-Gaussian which is likely to account for the difference.  Fig.~\ref{fig:vis} shows the CGPS sampled visibilities compared to the AMI visibilities, and the flux density ratio derived from fitting to the maps.  It can be seen that the correlation between the visibilities is very good, indicating a good morphological correlation between 1.42 and 15\,GHz and supporting the rising spectrum derived by using the CGPS data to correct the AMI flux densities for flux loss.

\begin{figure}
  \begin{center}
	\centerline{
%
%
\begin{psfrags}%
\psfragscanon%
\Large
\psfrag{s08}[t][t]{\color[rgb]{0,0,0}\setlength{\tabcolsep}{0pt}\begin{tabular}{c}Frequency (GHz)\end{tabular}}%
\psfrag{s09}[b][b]{\color[rgb]{0,0,0}\setlength{\tabcolsep}{0pt}\begin{tabular}{c}Flux density (Jy)\end{tabular}}%
%
\psfrag{x01}[t][t]{0}%
\psfrag{x02}[t][t]{0.1}%
\psfrag{x03}[t][t]{0.2}%
\psfrag{x04}[t][t]{0.3}%
\psfrag{x05}[t][t]{0.4}%
\psfrag{x06}[t][t]{0.5}%
\psfrag{x07}[t][t]{0.6}%
\psfrag{x08}[t][t]{0.7}%
\psfrag{x09}[t][t]{0.8}%
\psfrag{x10}[t][t]{0.9}%
\psfrag{x11}[t][t]{1}%
\psfrag{x12}[t][t]{17}%
\psfrag{x13}[t][t]{15}%
\psfrag{x14}[t][t]{0.5}%
\psfrag{x15}[t][t]{1}%
\psfrag{x16}[t][t]{5}%
\psfrag{x17}[t][t]{10}%
%
\psfrag{v01}[r][r]{0}%
\psfrag{v02}[r][r]{0.1}%
\psfrag{v03}[r][r]{0.2}%
\psfrag{v04}[r][r]{0.3}%
\psfrag{v05}[r][r]{0.4}%
\psfrag{v06}[r][r]{0.5}%
\psfrag{v07}[r][r]{0.6}%
\psfrag{v08}[r][r]{0.7}%
\psfrag{v09}[r][r]{0.8}%
\psfrag{v10}[r][r]{0.9}%
\psfrag{v11}[r][r]{1}%
\psfrag{v12}[r][r]{0.22}%
\psfrag{v13}[r][r]{0.24}%
\psfrag{v14}[r][r]{0.26}%
\psfrag{v15}[r][r]{0.28}%
\psfrag{v16}[r][r]{0.1}%
\psfrag{v17}[r][r]{0.2}%
\psfrag{v18}[r][r]{0.3}%
\psfrag{v19}[r][r]{0.4}%
\psfrag{v20}[r][r]{0.5}%
%
\resizebox{0.41\textwidth}{!}{\includegraphics{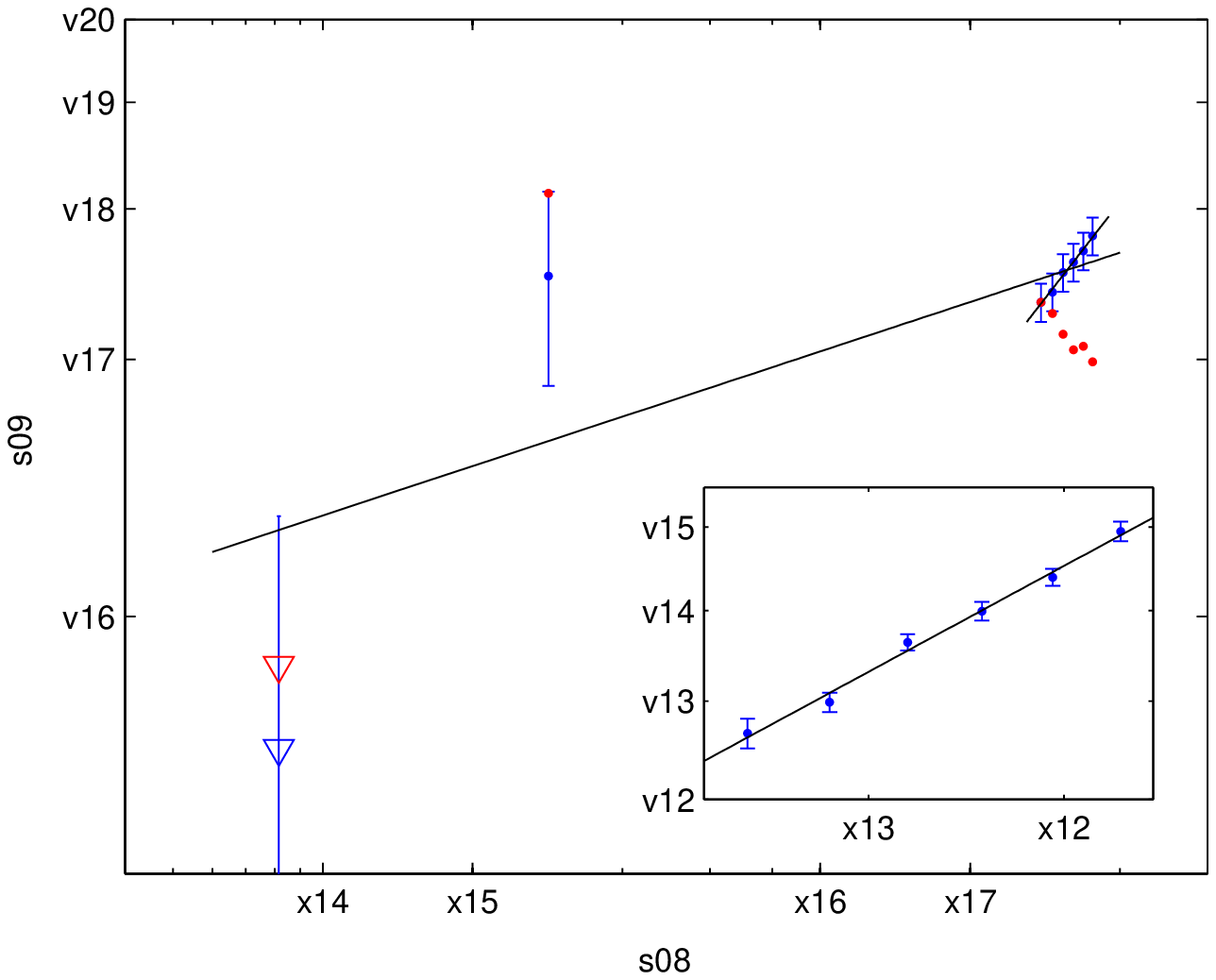}}%
\end{psfrags}%
%
\qquad\includegraphics[width=0.45\textwidth]{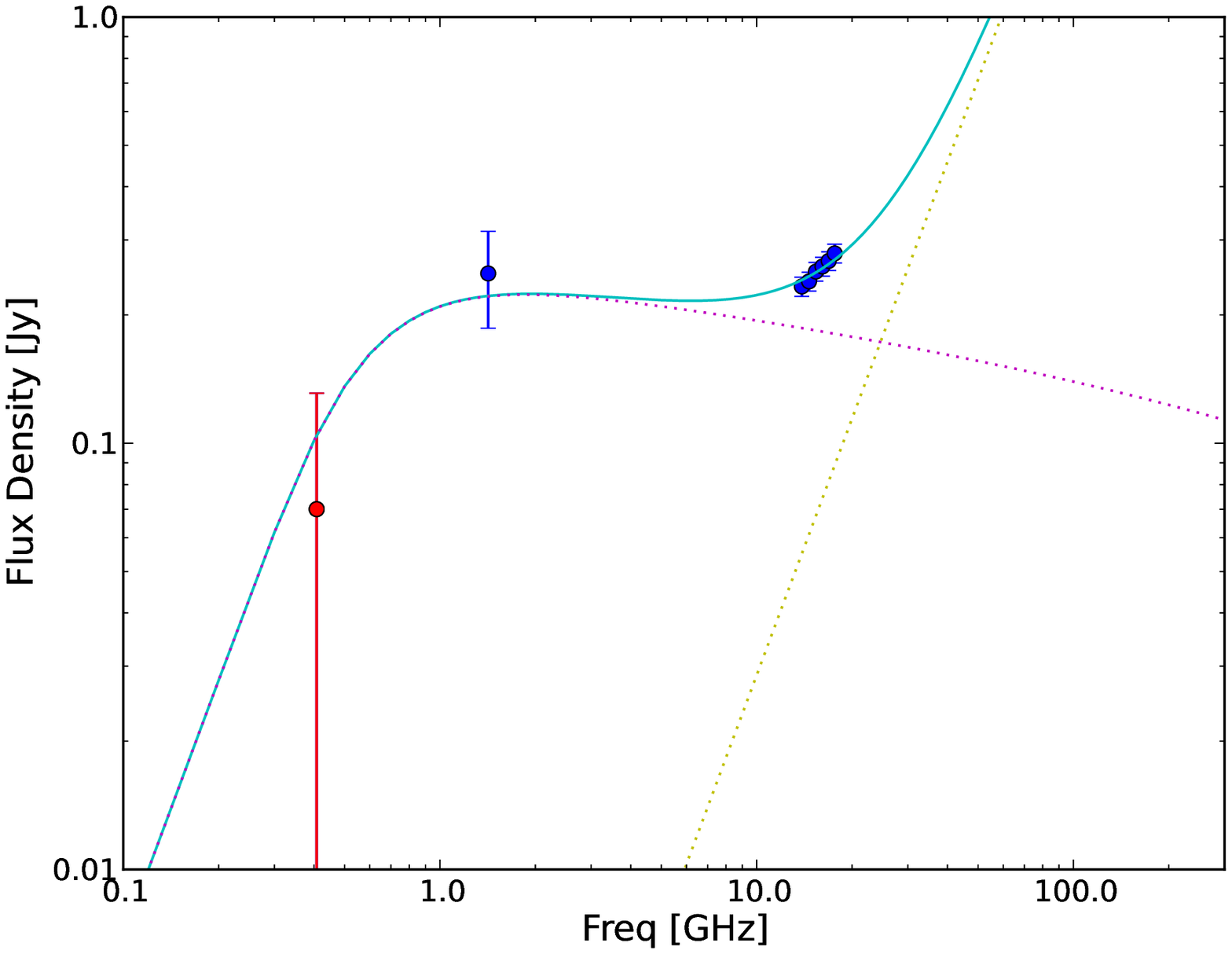}}
	\centerline{(a) \hskip .45\textwidth (b)}
    	\caption{SED for AME-G107.1$+$5.2.  (a) shows the fluxes at 0.408 and 1.42\,GHz from the CGPSE catalogue normalised by the flux loss expected in AMI channel 3, and the flux-loss-corrected AMI fluxes.  The markers and spectral index fits shown are as in Fig.~\ref{Fi:PSJ0541_spec}. (b) shows a fit to the flux densities using a two-component model consisting of a classical and a hyper-compact \textsc{Hii} region.\label{Fi:PSJ2219_spec}}
  \end{center}
\end{figure}

\begin{figure}[ht]
  \begin{center}
    \includegraphics[bb=46 138 570 668, clip=,width=0.8\textwidth]{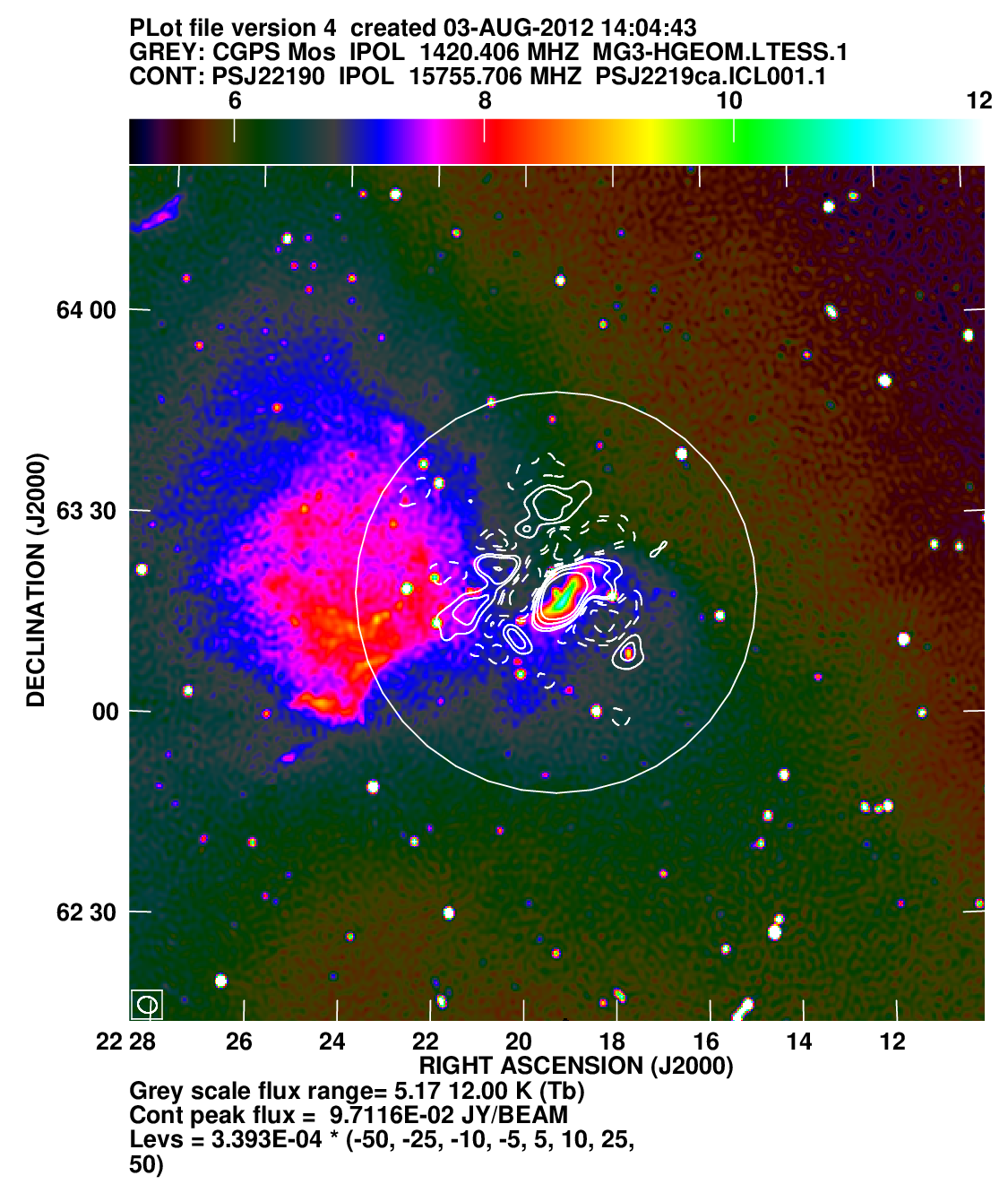}
    \caption{\textbf{AME-G107.1$+$5.2:} AMI SA data is shown as contours over CGPS 1.4\,GHz pseudo-colour, in brightness temperature.  The pseudo-colour scale is truncated at 12\,K ($\simeq$ the peak brightness of S140) to show the low-surface-brightness diffuse structures.  AMI contours are at $\pm$5, 10, 25, and 50 $\sigma_{rms}$ on the AMI map (0.34\,mJy\,beam$^{-1}$).  The 1$^{\circ}$ \emph{Planck} smoothing scale is shown as the large circle and the AMI SA synthesised beam is shown in the bottom left corner. \label{Fi:PSJ2219}}
  \end{center}
\end{figure}

Fitting to the sampled CGPS 1.42\,GHz flux density and AMI channel 3 \textit{uv}-coverage gives a spectral index of $\alpha=-0.02\pm0.2$, consistent with optically thin free-free emission.  Fluxes at lower resolution were not estimated due to the more complex background of this source; the CGPS 408\,MHz catalogue flux density adjusted by the same flux loss percentage as the 1.42\,GHz value is included as a reference however the errors are extremely large.  Fig.~\ref{Fi:PSJ2219_spec}(a) shows the SED; it seems likely that the AMI fluxes indicate the presence of AME or an ultra/hyper-compact \textsc{Hii} region.  Fig.~\ref{Fi:PSJ2219} shows the AMI SA channel-averaged map and the corresponding CGPS 1.4\,GHz map. 

Two scenarios are possible that may explain the rising spectrum across the AMI band: scenario 1 is the presence of a hyper-compact {\sc Hii} ({\sc HCHii}) region, and scenario 2 is the presence of a spinning dust component. Scenario 1 is illustrated in Fig.~\ref{Fi:PSJ2219_spec}(b) where the lower frequency data is dominated by free-free emission from plasma with an emission measure of $EM = 5.26\pm1.5\times 10^5$\,cm$^{-6}$\,pc assuming an electron temperature of $T_{\rm e} = 8000$\,K. This would indicate a region of classical {\sc Hii}, however the relatively high emission measure implied by the turnover frequency lying between 408\,MHz and 1.4\,GHz requires that the source have a much smaller angular size ($\sim0.1$\,arcmin) than that fitted to the AMI map ($\Omega_{\rm S140} = 242.5\times 109.3''$ P.A. = 148.7$^{\circ}$) to be consistent with the magnitude of the flux density. In addition the best fitting emission measure for the high frequency component is $EM=5.37\times 10^11$\,cm$^{-6}$\,pc, which is extremely high. 

\begin{figure}
  \begin{center}
	\centerline{\includegraphics[width=0.45\textwidth]{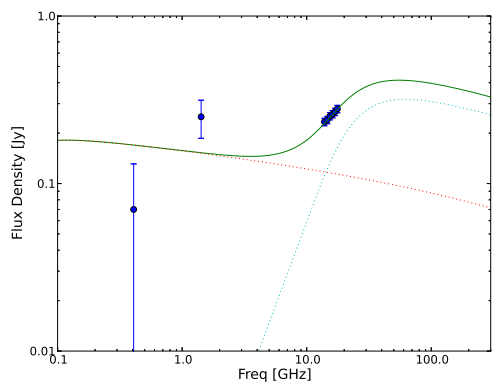}}
    	\caption{SED for AME-G107.1$+$5.2.  The fluxes at 0.408 and 1.42\,GHz are from the CGPSE catalogue normalised by the flux loss expected in AMI channel 3; the higher frequency fluxes are the flux-loss-corrected AMI fluxes.  Two free-free SED components are shown using a source size for the low frequency component fixed to that measured from the AMI-SA maps.\label{fig:s140}}
  \end{center}
\end{figure}

Fixing the source size to that measured from the AMI data produces a poorer fit to the low frequency data, with an emission measure of $EM=2.05\pm0.41\times 10^3$\,cm$^{-6}$\,pc, see Fig.~\ref{fig:s140}. The rising spectrum across the AMI band requires a second free-free component with $EM>7\times 10^8$\,cm$^{-6}$\,pc, consistent with an ultra- or hyper-compact {\sc Hii} region. In this case we find a best fitting emission measure for the high frequency component of $2.16\times 10^9$\,cm$^{-6}$\,pc with a size of $\sim 0.001$\,pc consistent with an object on the ultra/hyper compact border. In spite of the poorer fit to the low frequency data we consider this case more likely than the first as it is consistent with the measured source extent and predicts emission measures which are more physically feasible.

\begin{figure}
  \begin{center}
	\centerline{\includegraphics[width=0.45\textwidth]{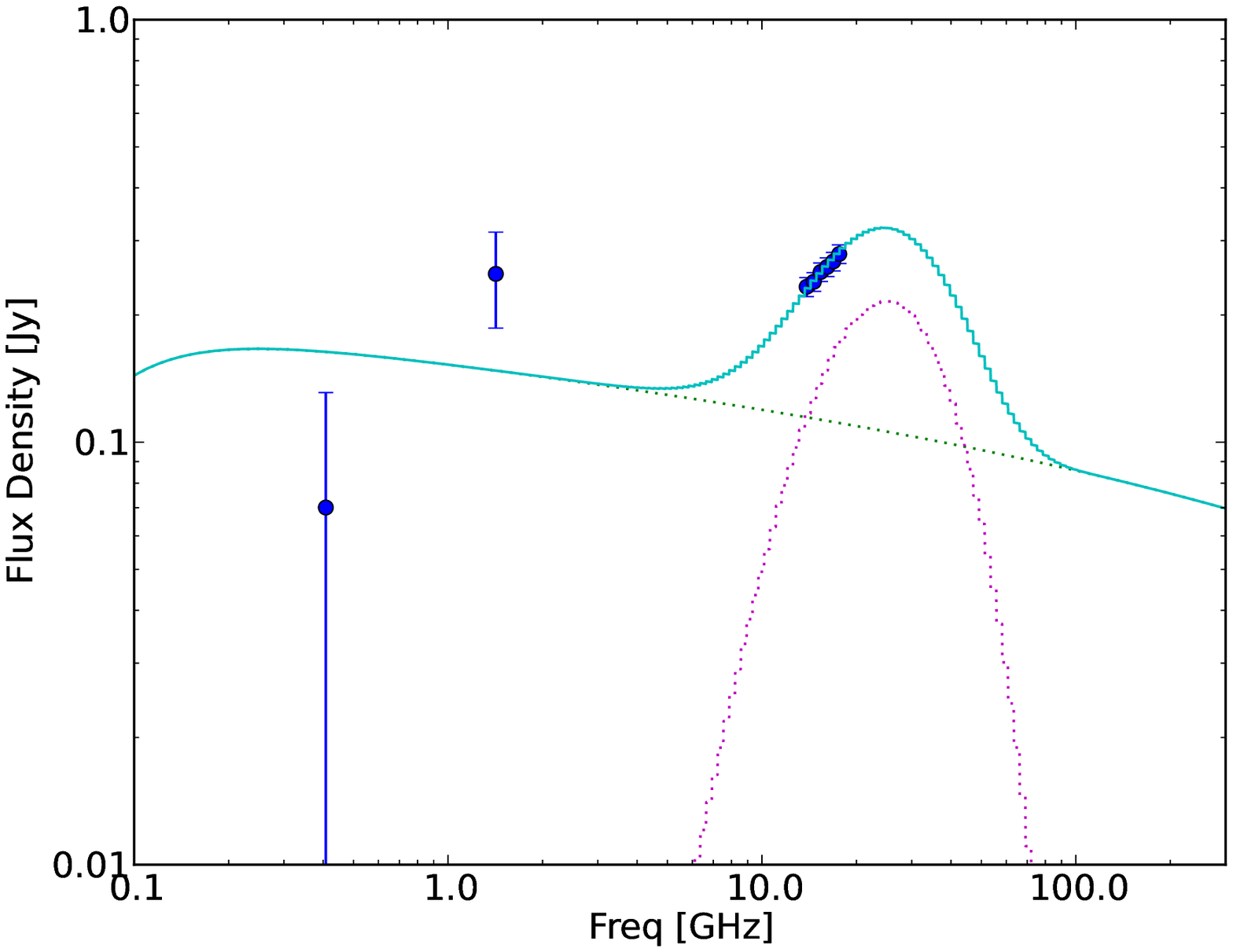}\qquad\includegraphics[width=0.45\textwidth]{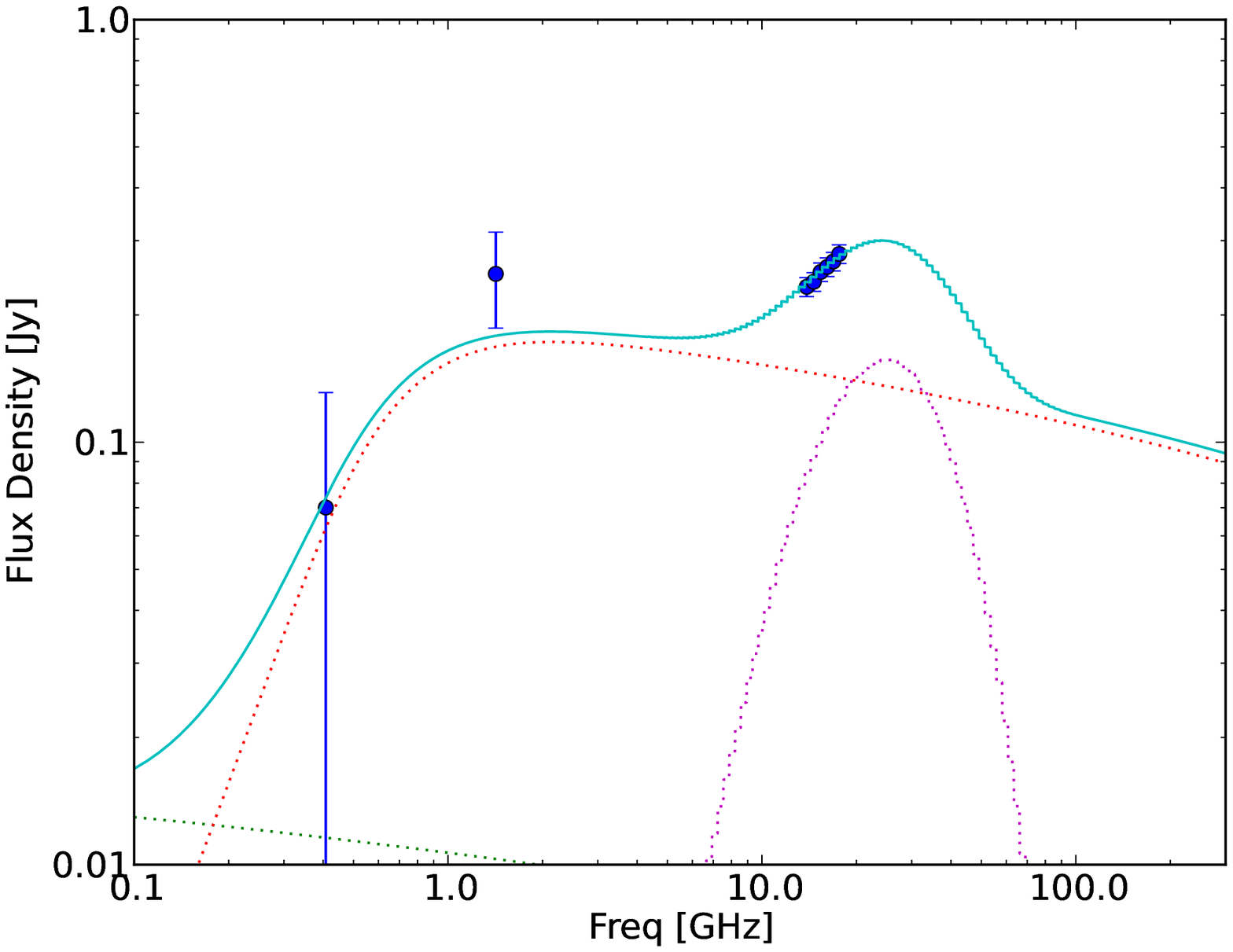}}
	\centerline{(a) \hskip .45\textwidth (b)}
    	\caption{SED for AME-G107.1$+$5.2 including fluxes at 0.408 and 1.42\,GHz from the CGPSE catalogue normalised by the flux loss expected in AMI channel 3, and the flux-loss-corrected AMI fluxes. (a) shows a single free-free component plus spinning dust contribution to the SED; (b) shows a fit to the flux densities using a two-component free-free model plus a spinning dust contribution. Please see the text for details. \label{fig:ffspff}}
  \end{center}
\end{figure}

Scenario 2 is illustrated in Fig.~\ref{fig:ffspff}(a) where the rising spectrum in the AMI data is accounted for by the introduction of a spinning dust model, here taken as the Draine \& Lazarian Warm Ionized Medium (WIM) model \cite{1998ApJ...494L..19D}. For the small angular size of the emitting region the column density of the spinning dust emitting region needs to be reasonably high at $n_{\rm H}=6.14\pm1.09~\times 10^{22}$\,cm$^{-2}$. A further possibility is that the 408\,MHz datum defines an optically thin free-free component, whilst the 1.4\,GHz point represents a contribution from a more compact region of {\sc Hii} and the rising spectrum across the AMI band indicates a \emph{further} component arising from even more compact {\sc Hii} or spinning dust. This possibility is shown in Fig.~\ref{fig:ffspff}(b); interestingly the optically thin free-free component shown has a similar emission measure to that fitted to the low frequency data on larger scales by \cite{2011A&A...536A..20P}, although the uncertainty on the fitted emission measure due to the large error on the 408\,MHz point makes the comparison weak.

With only the up-turn in the SED visible over the AMI band it is not possible to distinguish between these two scenarios using the AMI data alone. We do however confirm that a rising spectrum is present at frequencies above 10\,GHz. However, we also note that the magnitude of the emission required to explain the rising AMI spectrum is approximately an order of magnitude too small to explain the excess observed with \emph{Planck}.

\section{Discussion}

Detections of AME from \emph{Planck} data are based on flux densities integrated within an aperture of 1-degree radius. Such measurements will contain contributions from a variety of discrete sources, in addition to the large-scale diffuse background. The data presented here allow us to examine the population of discrete sources that contribute to these detections and investigate the physics behind their emission over a frequency range sensitive to the proposed spinning dust mechanism, which is generally considered to provide the physical basis for AME. 

The data presented here have shown that the emission from S235, the dominant source within the G173.6+2.8 region, has a spectrum consistent with optically thin free-free emission and shows no indication of a spinning dust component. For the source S140, dominant within the G107.1+5.2 region, the data presented here have shown that the spectrum of this source rises across the AMI band and is not consistent with an extrapolation of optically thin free-free from lower frequencies.

\subsection{The spectrum of G107.1+5.2}

\begin{figure}
\centerline{\includegraphics[width=0.45\textwidth]{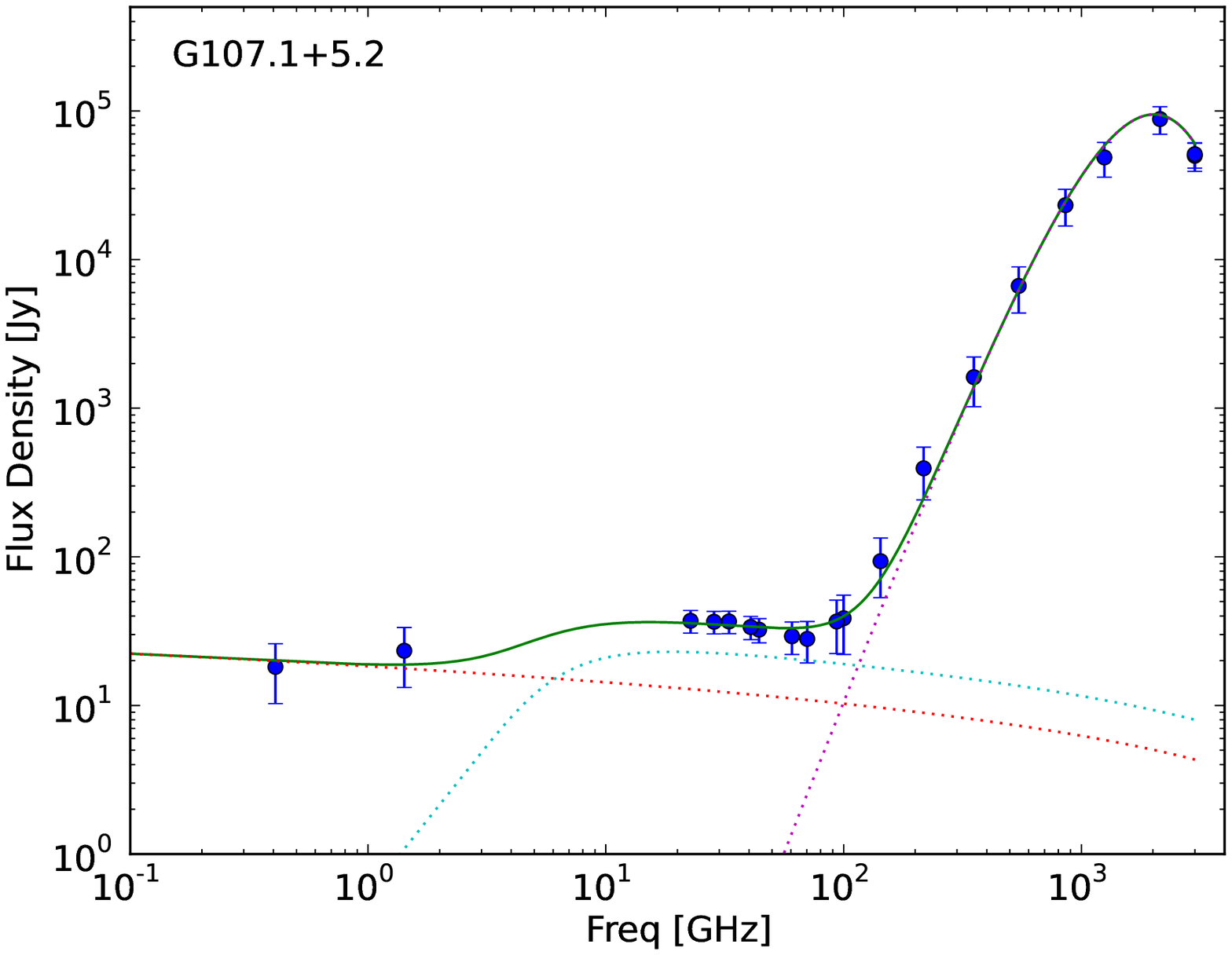}\qquad\includegraphics[width=0.45\textwidth]{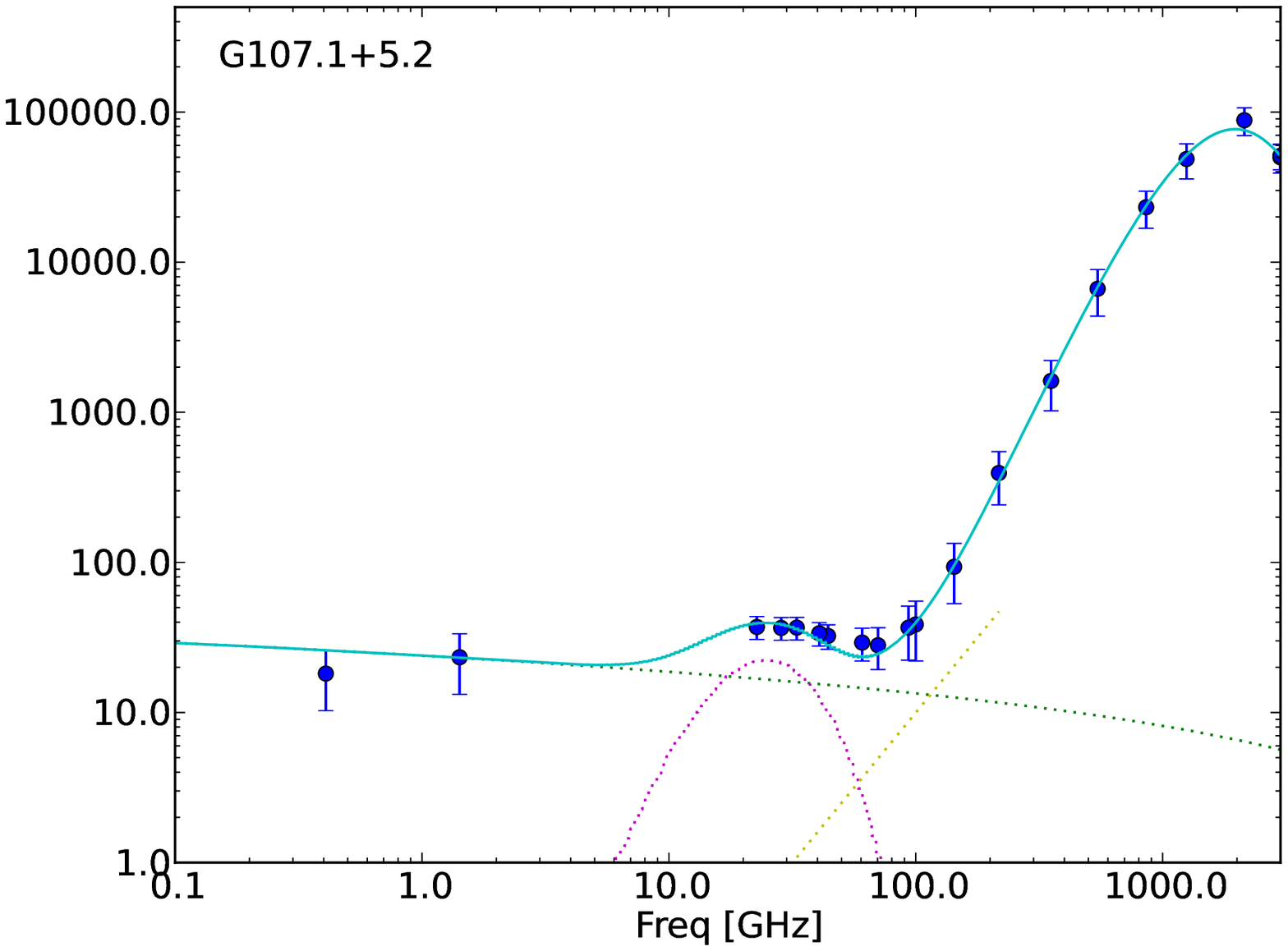}}
\centerline{(a)\hskip 0.45\textwidth (b)}
\caption{(a) G107.1+5.2 fitted with a two component free-free model. The two free-free components are shown as red and blue dotted lines, and the contribution from the thermal dust greybody as a purple dotted line. The combined model is shown as a solid green line. Model parameters are listed in Table~\ref{tab:ff}. (b) G107.1+5.2 fitted with a free-free plus spinning dust model. The free-free component is shown as a green dotted line and the spinning dust component as a purple dotted line. The Rayleigh-Jeans tail of a CMB component is shown as a yellow dotted line, with the thermal dust greybody model excluded for clarity. The combined model is shown as a blue solid line. Data points for both (a) and (b) are as in \cite{2011A&A...536A..20P}. \label{fig:sed2}}
\end{figure}

On large scales the spectrum of G107.1+5.2 \cite{2011A&A...536A..20P} is ambiguous in nature. Here we examine that ambiguity by comparing different physical scenarios through their spectra (data points from C. Dickinson, priv. comm.). In Fig.~\ref{fig:sed2}(a) we show a scenario where the excess emission at microwave frequencies is provided by the presence of an ultra-compact {\sc Hii} region and in Fig.~\ref{fig:sed2}(b) we show a scenario where the excess emission at microwave frequencies is provided by a spinning dust component. In both cases maximum likelihood (ML) fitted models are shown. These fits were found using an MCMC based method which calculates both ML parameters with associated uncertainties, as well as the evidence values for the specific model using a simulated annealing approach \cite{2004ApOpt..43.2651H}. 

We consider models of the same form as the original \emph{Planck} analysis with the electron temperature of the free-free emitting gas fixed to $T_{\rm e} = 8000$\,K, but leave the emission measure free to vary. Similarly we fix the temperature of the dust component to be $T_{\rm d} = 18.96$\,K \cite{2011A&A...536A..20P} as this is determined largely by the peak of the dust greybody; we retain the opacity index, $\beta$, as a free parameter as well as a normalization co-efficient. To these two initial components we add a second free-free component, parameterized by a second emission measure and by a size which is defined relative to the size of the first free-free emitting region. The electron temperature of this second component is also fixed at $T_{\rm e} = 8000$\,K.

\begin{table}
\caption{Maximum likelihood parameters for two component free-free model fitted to G107.1+5.2.\label{tab:ff}}
\begin{center}
\begin{tabular}{lccc}
\hline \hline
Parameter & {\sc Hii} & {\sc UC/HCHii}$^{\ddagger}$ & Thermal Dust \\
\hline
$T_{\rm e}$ [K] & 8000 & 8000 & $-$ \\
$T_{\rm d}$ [K] & $-$ & $-$ & 18.96 \\
$EM$ [cm$^{-6}$\,pc] & $147\pm47$ & $4.4\times 10^7$  & $-$ \\
$\beta$ & $-$ & $-$ & $2.20\pm0.18$ \\
size$^{\dagger}$ [arcmin] & $60$ & $0.16$ & $60$ \\
\hline
\end{tabular}
\begin{minipage}{0.7\textwidth}
$^{\dagger}$ radius of circular aperture.
$^{\ddagger}$ listed parameters represent best fit values, uncertainties are highly non-symmetric.
\end{minipage}
\end{center}
\end{table}

\begin{table}
\caption{Maximum likelihood parameters for spinning dust model fitted to G107.1+5.2.\label{tab:sp}}
\begin{center}
\begin{tabular}{lcccc}
\hline \hline
Parameter & {\sc Hii} & Spinning Dust & Thermal Dust & CMB \\
\hline
$T_{\rm e}$ [K] & 8000 & $-$ & $-$ & $-$\\
$T_{\rm d}$ [K] & $-$ & $-$ & 18.96 & $-$\\
$EM$ [cm$^{-6}$\,pc] & $235\pm48$ & $-$  & $-$ & $-$\\
$\beta$ & $-$ & $-$ & $1.91\pm0.11$ & $-$ \\
size$^{\dagger}$ [arcmin] & $60$ & $60$ & $60$ & $60$\\
$n_{\rm H}$ [$10^{21}$\,cm$^{-2}$] & $-$ & $4.59\pm1.13$ & $-$ & $-$ \\
$\Delta T$ [$\mu$K] & $-$ & $-$ & $-$ & $34\pm4$ \\
\hline
\end{tabular}
\begin{minipage}{0.7\textwidth}
$^{\dagger}$ radius of circular aperture.
\end{minipage}
\end{center}
\end{table}

In the second scenario we assume an identical first free-free component, and similarly a dust greybody with a fixed dust temperature. To these two components we add a spinning dust component parameterized by a column density, $n_{\rm H}$, and include a potential CMB contribution in a similar fashion to \cite{2011A&A...536A..20P}. The spinning dust component that we use is the Draine \& Lazarian (1998) WIM model. The model parameters we obtain differ slightly from those of \cite{2011A&A...536A..20P} but are consistent within errors

In the first scenario we find that the data are well fitted by a two component free-free model where the second component has an emission measure of $EM>10^7$\,cm$^{-6}$\,pc and a source size which is greater than a factor of $\sim$1200 smaller than the size of the first emitting region. Both constraints are consistent with the presence of an ultra-compact {\sc Hii} region as the relative source size indicates an object of $<0.1$\,pc at a distance of 910\,pc. The lack of data in the frequency range 1.4-20\,GHz makes these two parameters highly degenerate as the optical depth of such an object can reach unity anywhere over an order of magnitude in frequency. Indeed a rising spectrum over the AMI band, as observed, would indicate a turn-over frequency at $\nu_{\rm t} >15$\,GHz and consequently an emission measure of $EM>7\times 10^8$\,cm$^{-6}$\,pc, consistent with the fitted constraints.

In the second scenario we find parameter values similar to those determined by \cite{2011A&A...536A..20P}, with the exception of the CMB contribution where we find a maximum likelihood value of approximately half that determined by \cite{2011A&A...536A..20P}. Maximum likelihood parameters for both models are listed in Tables~\ref{tab:ff}~\&~\ref{tab:sp}. These models are shown fitted to the data in Fig.~\ref{fig:sed2}(a) \& (b). The logarithmic difference in the evidence between the two models, as determined over a 3\,$\sigma$ prior volume, is $\Delta \ln Z = 36.0\pm0.7$ in favour of the spinning dust model. 

Assuming that the optically thin free-free component which dominates the \emph{Planck} SED arises on scales which are not visible to AMI, we examine the possibility that the rising spectrum observed across the AMI band and the excess emission detected by \emph{Planck} can both be explained by a single hyper-compact {\sc Hii} region. If this is the case, we predict that the emission seen across the AMI band in a \emph{Planck}-size aperture would be simply the sum of an extrapolation of the optically thin component to these frequencies plus the measured AMI flux densities. This is subject to a couple of strong assumptions: firstly, that the emission from the {\sc HCHII} region arises on scales which are compact to the AMI beam and there will consequently be no flux loss relative to the \emph{Planck} measurement; secondly, that the {\sc HCHii} is the dominant contribution to the AMI flux densities. The first of these assumptions is reliable as {\sc HCHii} exists on scales which are highly compact to both the AMI and \emph{Planck} beams. The second assumption is less reliable as it depends on the parameters of the {\sc HCHii} model, which are not possible to constrain absolutely from the AMI data alone. Considering this to be the case it is immediately obvious that the magnitude of the AMI flux densities is insufficient to be compatible with the excess seen by \emph{Planck} being contributed by a single {\sc HCHii} region. 

By the same token it is also not possible for the \emph{Planck} excess to be explained by the presence of a spinning dust region with angular scales entirely recovered by AMI. Since the possibility of an extended region of {\sc HCHii} has no precedent in the literature, it seems likely then that the excess measured by \emph{Planck} towards G107.1+5.2 arises from a spinning dust region, the majority of which is extended on scales larger than those measured by AMI. This conclusion is supported by the strong evidence in favour of a spinning dust component found by the model comparison.

\section{Conclusions}\label{conclusions}

In the case of both G173.6+2.8 (S235) and G107.1+5.2 (S140) we conclude that the bulk of the excess emission seen by \emph{Planck} must arise on scales larger than 10\,arcmin. In the case of G173.6+2.8 we confirm that the dominant source within the \emph{Planck} aperture, S235, has a spectrum on scales of $2-10$\,arcmin consistent with optically thin free-free emission. In the case of G107.1+5.2 we demonstrate that the dominant source, S140, has a rising spectrum across the AMI band. This spectrum is consistent with either spinning dust emission or the presence of {\sc UC/HCHii}. With only the low frequency data available on these scales it is not possible to distinguish between the two mechanisms, however, we conclude that the magnitude of the contribution causing the rising spectral index across the AMI band is not sufficient to explain the measured \emph{Planck} excess, which must once again arise on scales larger than those probed by AMI. In this case, it is therefore possible that the rising spectrum over the AMI band is also a consequence of spinning dust emission.

\section*{Acknowledgments}
We thank the staff of the Mullard Radio Astronomy Observatory for their invaluable assistance in the commissioning and operation of AMI, which is supported by Cambridge University and the STFC.  We also thank Clive Dickinson for the data points for the large-scale spectrum of G107.1+5.2.  This research has made use of the SIMBAD database, operated at CDS, Strasbourg, France.  YCP acknowledges the support of a CCT/Cavendish Laboratory studentship.

\bibliographystyle{abbrv}
\bibliography{main}

\end{document}